%% file: paley_paper3.tex
\documentclass[conference]{IEEEtran}
\usepackage{empheq}
\usepackage{amsmath}
\usepackage{amsthm}

\usepackage{mathtools}
\usepackage{ upgreek }

\PassOptionsToPackage{hyphens}{url}

\usepackage[hyphens]{url}

\usepackage{fullpage,subfigure}
\usepackage{graphicx,amsmath,amsfonts,amscd,amssymb,bm,cite,epsfig,epsf,color}
\usepackage[colorlinks=true,urlcolor=blue,citecolor=blue]{hyperref}
\usepackage[hyphenbreaks]{breakurl}
\hypersetup{breaklinks=true}
\usepackage{array} 
\usepackage[small,bf]{caption}
\usepackage{bbm}
\usepackage{pgfplots}
\usetikzlibrary{spy}
\usetikzlibrary{arrows,automata}
\usepackage{blkarray}

\newtheorem*{lemma*}{Lemma}

\title{Revisiting Block-Diagonal SDP Relaxations for the Clique Number of the Paley Graphs}

\author{\IEEEauthorblockN{Vladimir A. Kobzar}
\IEEEauthorblockA{Department of Applied Physics
and \\ Applied Mathematics, Columbia University\\  
vak2116@columbia.edu}
\and
\IEEEauthorblockN{ Krishnan Mody}
\IEEEauthorblockA{Courant Institute of Mathematical Sciences\\ New York University, New York, NY \\km2718@nyu.edu}} 
\input{macros}

\begin{document}

\maketitle

\begin{abstract} This work addresses the block-diagonal semidefinite program (SDP) relaxations for the clique number of the Paley graphs. The size of the maximal clique (clique number) of a graph is a classic NP-complete problem; a Paley graph is a deterministic graph where two vertices are connected if their difference is a quadratic residue (square) in a finite field with the number of elements given by certain primes and prime powers. Improving the upper bound for the Paley graph clique number for prime powers that are non-squares is an open problem in combinatorics. Moreover, since quadratic residues exhibit pseudorandom properties, Paley graphs are related to the construction of deterministic restricted isometries, an open problem in compressed sensing. Recent work provides numerical evidence that the current upper bounds can be improved by the sum-of-squares (SOS) relaxations. In particular, the bounds given by the SOS relaxations of degree 4 (SOS-4) have been empirically observed to be growing at an order smaller than square root of the prime. However, computations of SOS-4 appear to be intractable with respect to large graphs. Gvozdenovic et al. introduced a more computationally efficient block-diagonal hierarchy of SDPs and computed the values of these SDPs of degrees 2 (L2) for the Paley graph clique numbers associated with primes p less or equal to 809, which bound from above the corresponding SOS-4 relaxations. We compute the values of the L2 relaxations for p's between 821 and 997. Our results provide some numerical evidence that these relaxations, and therefore also the SOS-4 relaxations, may be scaling at an order smaller than the square root of p. However, due to the size of the SDPs, we have not been able to compute L2 relaxations for p's greater than 997. Therefore, our scaling estimate is not conclusive and presents an interesting open problem for further study.  
\end{abstract} 

\section{The clique number of the Paley graphs}

A \emph{Paley graph} $G_q$ is a graph with  $q$ vertices, where $q$ is a prime power such that $q = 1 \mod 4$;\footnote{By the law of quadratic reciprocity, this condition   ensures that $-1$ is a quadratic residue in $\mathbbm F_q$. Therefore, if $i-j$ is a quadratic residue then so is $j-i$, and the graph is undirected.} two vertices are connected by an edge $\{i,j\}$ whenever $i-j $ is a quadratic residue in $\mathbbm F_q$. (We may sometimes refer to quadratic residues as \emph{squares}, and to nonresidues as \emph{nonsquares}.)\footnote{See  \cite{Jon20} for general background on Paley graphs.}

The \emph{clique number} $\omega(G)$ of a graph $G$ is the number of vertices in its largest complete subgraph or \emph{clique}. For any Paley graph $G_q$, this number is bounded above by $\sqrt {q}$ \cite {Yip22}. 

For a Paley graph with $q = p^{2k}$ where $k$ is a positive integer, the foregoing bound is tight \cite{BDR88}. However, less is known about  $\omega(G_q)$  when $q$ is a prime power that is a non-square.  In particular, if $q=p$, the state-of-the-art lower bounds are scaling as  $\log p \cdot \log \log \log p $; the  $\log \log \log p$ term can be improved to  $ \log \log p$ conditional on the Generalized Riemann Hypothesis (see \cite{GR90} and Theorem 13.5 in \cite{ Mon71}). On the other hand, the existing state-of-the-art upper bounds in references  \cite{HP21, DBSW21} improve on $ \sqrt  {p}$ only by a constant prefactor. We will refer to the upper bound $(\sqrt {2p-1}+1)/2$ in those references as $HP(G_p)$. Numerical experiments appear to suggest that the Paley graph clique number is polylogarithmic in $p$ (see discussion of \cite{Shearer96,Exo22} in \cite{BMR}). However, proving even a $p^{\frac{1}{2}-\epsilon}$  bound for some $\epsilon>0$ is regarded as a difficult open problem in additive combinatorics \cite{CL07}, sometimes referred to as the \emph{square root bottleneck} (see also \cite{kunisky2023spectral}). 

\section{Connections to deterministic restricted isometries}

Paley graphs are connected to the construction of deterministic $M \times N$ matrices  with the restricted isometry property (RIP),  an important problem in compressed sensing and sparse recovery \cite{Tao08}.  Random matrix constructions achieve RIP when  sparsity is on the order of $M/ \text{polylog} (N)$. However, most deterministic constructions, such as equiangular frames (\emph{ETFs}), are based on controlling a certain \emph{coherence} value, which achieves RIP only when sparsity is on the order of $\sqrt {M}$; this limitation is known as the \emph{square root bottleneck}.\footnote{Reference \cite{ABSM15} designed deterministic RIP matrices that support the same sparsity $\sqrt {M}$ based on the coherence analysis; however, they are constructed using the adjacency matrix of a Paley graph rather than an ETF.}    The only unconditional construction that overcomes this bottleneck was provided in \cite {B+11a, B+11b}, which leveraged additive combinatorics techniques  to achieve RIP for $\Omega (M^{\frac{1}{2} +\epsilon})$ sparsity for small $\epsilon>0$ (see also \cite{Mix15}). 

Reference \cite{BFMW13} constructed a family of deterministic ETF matrices using the quadratic residues modulo a prime number $p$ (the \emph{Paley matrices}) which  provably achieve RIP when sparsity is on the order of $\sqrt {p} $ by the aforementioned coherence analysis but are \emph{conjectured} to achieve it when sparsity on the order of $p/ \text{polylog}(p)$ (which would match the random construction if $p$ is proportional to $M$).  Reference  \cite{BFMM16}  used a matrix construction based on the Legendre symbol (which is closely connected to Paley graphs) to reduce the number of random bits in a random RIP matrix.  

Finally, conditioned on a conjecture about the number of edges in any subgraph of a Paley graph, the Paley matrices overcome the square root bottleneck\cite {BMJ17}.\footnote{Using a similar analysis, reference \cite {KPB19} showed an improvement on the \emph{square root bottleneck} by $\epsilon = \frac{9}{40}+ \kappa$ for small $\kappa$; while this result is not  conditioned on any conjectures, it only holds for signals with a certain sparse structure.} In this conditional construction,  a lower bound on $\omega(G_p)$  would lead to a lower bound on the distortion in the sparse recovery  (Theorem 2.3 in  \cite {BMJ17}). 

\section{SDP relaxations of the clique number}

The clique number  $\omega(G)$ of a graph $G$ is a classical NP-complete problem. It can be formulated as a polynomial optimization over $x \in \R^n$ where $n$ is the number of vertices of $G = (V,E)$:
\begin{equation*}
\label{clique_program}
\omega(G) = \left\{\begin{array}{ll}
    \max& \sum_{i \in V} x_i \\
    \text{s.t.} & x  \in \R^n, ~  x_i^2 = x_i \text{ for all } i \in V, \\
    & x_ix_j = 0 \text{ for all }\{ i, j\} \notin E \end{array}\right\}.
\end{equation*}
An extensive body of literature considered upper bounds produced by convex relaxations, which are more computationally efficient. One particular question in this literature is whether semidefinite program (SDP) relaxations would lead to an $O (n^{\frac{1}{2}-\epsilon})$ upper bound on the clique number for some $\epsilon>0$.

\emph{Erdos-Renyi  graphs} $G\sim \mathcal G(\frac{1}{2}, n)$ are random graphs where each edge is present independently with probability $\frac{1}{2}$. In this setting, reference \cite{FK03} showed that the  \emph{Lovasz-Schrijver hierarchy} of SDPs attains an $\Omega  (\sqrt {n})$ lower bound for clique number relaxations of any constant degree.   

Another line of work considered the \emph {sum of squares (SOS)} hierarchy of SDPs, also known as the \emph{Lasserre-Parrilo hierarchy}. In the context of the clique number problem, these relaxations, denoted by $SOS_{2t} (G)$ where $t$ is the degree of the hierarchy, are defined as follows.   Let $\mathcal{P}(V)$ be the collection of all subsets (power set) of $V$, and let $\mathcal{P}_{t} = \left\{ I \in \mathcal{P}(V) \,|\, |I| \leq t \right\}$ and $\mathcal{P}_{=t} = \left\{ I \in \mathcal{P}(V) \,|\, |I| = t \right\}$ denote the subsets of $V$ with at most $t$ and exactly $t$ elements, respectively.  For $y \in \mathbb{R}^{\mathcal{P}_{2t}}$, we define the \emph{moment matrix} of $y$ $M_t(y) \in \mathbb{R}^{\mathcal{P}_t \times \mathcal{P}_t}$ by $ M_t(y)_{IJ} = y_{I \cup J}$  where $I, J \in \mathcal{P}_t(V)$. We also denote by $\mathcal K$ the set of all cliques of $G$. Then the SOS hierarchy is given by
\begin{equation*}
SOS_{2t}(G)= \left\{\begin{array}{ll}
   \max & \sum_{i \in V} y_{i , \emptyset} \\
    \text{s.t.} & y \in \mathbb{R}^{\mathcal{P}_{2t}},  ~y_{\emptyset} = 1 \\
    &  y_{S, T} = 0 ~\forall ~S \cup T \notin \mathcal K\\
    & M_t(y)  \succeq 0
   \end{array}\right\}
    \end{equation*}
 In the average-case setting of  $\mathcal G(\frac{1}{2}, n)$, reference  \cite {BHK+19} established an $\Omega  (\sqrt {n})$ lower bound for the SOS relaxation of any constant degree for the clique number problem (see also earlier work \cite {DM15, RS15, HKP15} focusing on the $SOS_4$ relaxation). 

A \emph{stable set} of a graph $G = (V,E)$ is a subset $S\subset V$ such that no two nodes in $S$ have an edge between them. The size $\alpha(G)$ of the largest stable set of $G$ is called the independence or stability number of $G$.   The \emph{Lov\'{a}sz $\upvartheta$ function}, which can be also formulated as an SDP, is a convex relaxation of $\alpha(G)$. Note that for a complement graph $\bar G$, $\alpha(G)= \omega (\bar G) $, and Lov\'{a}sz $\upvartheta$   represents the first and weakest degree of the SOS hierarchy, i.e., $\upvartheta(G) = SOS_2(\bar{G}) $ (see, e.g., Section 4.1.3 in \cite {Gvozd}).  
 
 Since the Paley graphs are self-complementary, $\omega (G_p) = \alpha (G_p)$. The  classic upper bound $\alpha (G_p)\leq \sqrt {p}$ is realized by the $SOS_2(G_p)$ relaxation of the Paley graph clique problem (Theorem 13.14 in \cite{Bol01} and Theorem 8 in \cite{Lov79}), but the current state-of-the-art upper bound $HP(G_p)$ in \cite{HP21, DBSW21} has a tighter constant prefactor.   Numerical experiments in \cite {MMP19} show that the Lovasz  $\upvartheta$ relaxation with respect to appropriately chosen local subgraphs of $G_p$ with additional  Schrijver’s nonnegativity constraints often improves on  $HP(G_p)$; see also \cite{kunisky2023spectral}. 

Recent work revisited the higher-degree SOS relaxations in the deterministic context of the Paley graphs. Specifically,  for $q=p$ prime, reference \cite{KY22} presented numerical experiments suggesting that  $SOS_4(G_p)$  may scale  as $O(p^{\frac{1}{2}-\epsilon})$ for some $\epsilon>0$  and proved that these values are at least $\Omega  (p^{\frac{1}{3}})$. 

For large graphs, the $SOS$ relaxations appear to be computationally intractable, especially for higher degree $t$ of the hierarchy,  which entail optimization over $ \mathbb{R}^{\mathcal{P}_t \times \mathcal{P}_t}$ matrices. For example, the  $SOS_4$ relaxations do not appear to be currently computationally feasible for $p>250$ (Section 6 and Figure 1 in  \cite{KY22}).

\section{Block-diagonal SDP hierarchy}

References \cite{GLV, Gvozd}  introduced a new hierarchy of SDPs, denoted by $L^t$, which is nested between the Lovasz-Schrijver and SOS  hierarchies, and in particular the optimal values of this new hierarchy $L^t$ bound from above the corresponding SOS-$2t$ values. This new hierarchy is more computationally tractable that the SOS hierarchy because it is based on the block-diagonal submatrices of the moment matrix. In the context of the Paley graph clique number, the size of the block-diagonal relaxations can be reduced further by leveraging graph symmetries.  This section provides an exposition of the $L^2$ relaxations in \cite{GLV, Gvozd}.

For $y \in \mathbb{R}^{\mathcal{P}_{t+1}}$ and a subset of vertices $T \subset V$ of size $|T| = t-1$, let $M(T; y) \in \mathbb{R}^{\mathcal{P}_{t-1} \times (n+1)}$  be a principal submatrix of $M_t(y)$ whose rows and columns are indexed by $\mathcal{A}(T) = \bigcup_{S\subseteq T} \mathcal{A}_S$ where $\mathcal{A}_S = \{S\} \cup \left\{ S \cup \{i\} \,|\, i \in V \right\}$. Following  \cite{GLV}, we consider $\mathcal{A}_S$ as a multiset, i.e., we keep possible repeated occurrences, e.g.,  $S$ and $S \cup \{i\}$ if $i \in S$.\footnote{Therefore, $|\mathcal{A}_S| =2^{t-1} (n+1)$ and  $M(T; y) \in \mathbb{R}^{2^{t-1} (n+1) \times  2^{t-1} (n+1)}$. Technically $M(T; y)$ is a submatrix of  $M_t(y)$ after removing the repeating rows of the latter.}
Let $A_S(y)$ denote the principal submatrix of $M(T;y)$ indexed by the set $\mathcal{A}_S$:  it is a symmetric $(n+1)\times (n+1)$ matrix with entries
\begin{align*}
A_S(y)_{00} = y_S, A_S(y)_{0i}=y_{S\cup\{i\}}, ~ A_S(y)_{ij} = y_{S \cup \{i,j\}} 
\end{align*}
for $i, j \in V$. We will index matrix and vector entries by $[0,\dots,n-1]$, the elements of a finite field $\mathbb F_n$ to simplify our subsequent discussion  of the Paley graph  relaxations.  By Lemma 2.2 in \cite{GLV} $M(T;y)$ is positive semidefinite (PSD) if and only if for all $S\subseteq T$ the matrix 
		\[
			A(S,T)(y) := \sum_{S': S\subseteq S' \subseteq T} (-1)^{\left|S'\backslash S\right|}A_{S'}(y)
		\]
is PSD. 

These reductions lead to the following  relaxation of the independence set problem of an arbitrary graph with $n$ vertices:\footnote{The independence set formulation leads more naturally than the equivalent clique number formulation to further simplifications of the $L^t$ program discussed here in the context of Paley graphs.} $L^{t}(G)= $
\begin{subequations}
\begin{empheq}[left =  \empheqlbrace,  right=\empheqrbrace] {align*}
   \max & \sum_{i \in V} y_{\{i\} } \\
    \text{s.t.} & y \in \mathbb{R}^{\mathcal{P}_{t+1}},  ~y_{\emptyset} = 1 \\
    &  y_{\{i, j\}} = 0 ~\forall ~(i,j) \in E \\
    & A(S,T)(y)  \succeq 0 ~\text {for all}~ S \subset T ~ \text{and}~ T \in \mathcal{P}_{=t-1}
    \end{empheq}
    \end{subequations}
This optimization problem has $\binom{n}{t-1} 2^{t-1}$ PSD constraints  with respect to $n+ 1 \times n+1$ matrices $A(S,T)(y)$. 

In the remainder of this section, for consistency with \cite{GLV, Gvozd}, we will consider the independent set problem rather than the equivalent maximal clique problem.
		
In the context of a Paley graph $G_p$,  references \cite{GLV, Gvozd} exploit its symmetries to reduce the number of the PSD constraints. Using the vertex-transitivity of the graph and Lemma 2.4.5 in \cite{Gvozd}, the PSD constraints in $L^2(G_p)$ is reformulated in terms of two $(p+1) \times (p+1)$ matrices using just one arbitrary vertex $h \in V$, e.g., $\{0\}$; the resulting matrices $A_{\emptyset}(y)$ and $A_{\{0\}}(y)$ used in the constraints are given below.  Furthermore, a Paley graph is edge-transitive - this symmetry allows to reduce the number of optimization variables corresponding to non-edges to one variable,  $y_{\{0,k\}}$ for some ${\{0,k\}} \notin E$.  

The cliques and independent sets of size 3 (triangles) form orbits under the graph automorphism group of affine mappings $\phi_{ab}: \mathbbm F_p \rightarrow \mathbbm F_p$ given by $\phi_{ab}(u) = au+b$ where $u \in \mathbbm F_p$, $a,b \in \mathbbm F_p$ and $a \neq 0$ is a square in $\mathbbm F_p$. 
Therefore, the number of optimization variables corresponding to triangles without edges can be reduced to the number of the orbits of $\phi_{ab}$  acting on such triangles.  
 
Since a Paley graph is edge-transitive, the representatives of the orbits of fully connected triangles are given by $\{0, 1, \beta\}$ where both $\beta$ and $\beta-1$ are squares in $\mathbbm F_p$.  The representatives of the orbits of  $\phi_{ab}$ acting on triangles  without  any edges, which can be represented as $\{0, \alpha, \beta\}$, can be expressly computed as well.  Lemma 6.2.1 in   \cite{Passuello} explicitly sets forth their orbits; there are approximately $(p-5)/24$ orbits.  Let $\Omega $ be the set  of representatives of such orbits, and denote $m:=|\Omega|$.   Then
\begin{subequations}
\begin{empheq}[left = {L_{2}(G_p)=}  \empheqlbrace,  right=\empheqrbrace] {align}
   \max &\ p \cdot y_{\{0 \} } \\
    \text{s.t.} &\ y_{\{0\} },  y_{\{0, k\} } \in \R,~  y \in \mathbb{R}^{m} \\
      & A_{\emptyset}(y) -A_{\{0\}}(y)   \succeq 0  \label{eq:constr2}\\
        & A_{\{0\}}(y)  \succeq 0 \label{eq:constr1}
    \end{empheq}
    \end{subequations}
where  
\[
A_{\emptyset}(y) =   \left(\begin{array}{@{}c|c@{}}
    1 &  y_{\{0\} }  \mathbbm 1^\intercal  \\ \hline
     y_{\{0\} }  \mathbbm 1 & y_{\{0\} } I+   y_{\{0, k\} } A_{\bar{G}_p}
  \end{array}\right)
\]
encodes the vertices by $y_{\{0\} }$ and nonedges by $ y_{\{0,k\} } $, both scalars, and $A_ {\bar G_p}$ is the adjacency matrix of the complement graph $\bar {G}_p$.  Also we have  
\[
 A_{\{0\}}(y) =\left(\begin{array}{@{}c|c|c@{}}
    y_{\{0\} }  &y_{\{0\} } &  y_{\{0,k\} } q^\intercal \\ \hline
     y_{\{0\} }  & y_{\{0\} }&  y_{\{0,k\} } q^\intercal \\ \hline
     y_{\{0,k\} } q  &y_{\{0,k\} }q &   M\\
  \end{array}\right)
  \] 
 where the leftmost column $q := (A_{\bar{G}_p})_{1:\text{end}, 0}$ of the adjacency matrix $A_{\bar{G}_p}$, and
 \[
 M :=  y_{\{0,k\} }\text{diag} (q)+ \sum_{\{\alpha, \beta\} \in \Omega} y_{\{0, \alpha, \beta\} } X^{\alpha \beta}.
 \]
The matrix $X^{\alpha \beta} \in \R^{p-1\times p-1} $ encodes the  orbit of $\phi_{ab}$ acting on each representative triangle $\{0, \alpha, \beta\}$:   $X^{\alpha\beta}_{ij} = 1$  if  $\{i,j\} \in ~ \phi_{ab}(\{0, \alpha, \beta\})$ for any $a,b \in \mathbbm F_p$ where $a \neq 0$ is a square in $\mathbbm F_p$, and $X^{\alpha\beta}_{ij}=0$ otherwise.  (Note $X^{\alpha \beta}_{ii} =0$ for all $\{\alpha, \beta\} \in \Omega$.)

Since the second row and columns of $A_{\{0\}}(y)$ will match those of $A_{\emptyset}(y)$,  we can  remove the first row and column in each matrix for purposes of the  \eqref{eq:constr2} constraint.  This leads to a $p \times p$ rather than $p+1 \times p+1$ matrix in that constraint. Lastly for purposes of the other constraint \eqref{eq:constr1}, we can remove from  $A_{\{0\}}(y)$ the rows and columns with edges (which are indexed by the nonresidues)  leading to a $(p+1)/2 \times (p+1)/2$ matrix. 

\section{New Computations}
\begin{figure} 
\begin{center}
\includegraphics[scale=.62]{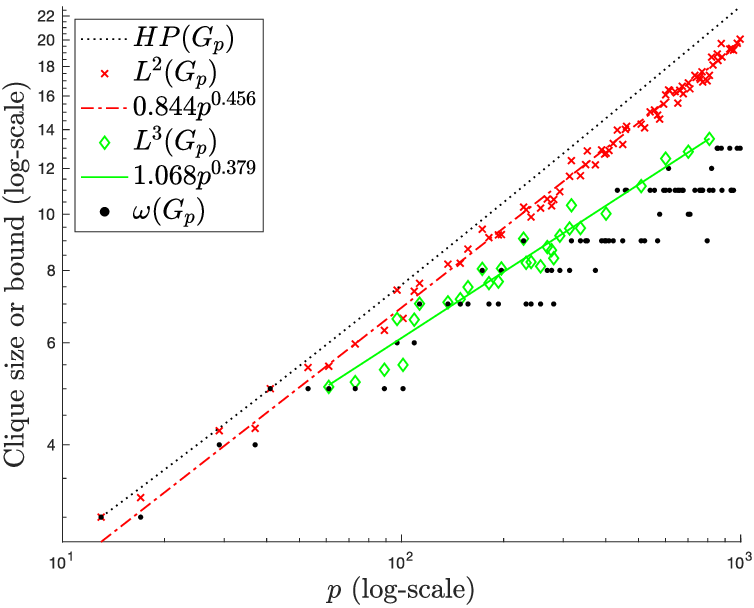}  
\end{center}
\caption{ The $L^2(G_p)$ values for $ 809< p < 1000$ determined in this paper, and the   $L^2(G_p)$  and $L^3(G_p)$ values for  $p \leq 809$ determined in \cite {GLV, Gvozd} are fitted to power models of the form $ap^b$. The values of $\omega(G_p)$ were obtained from \cite{Shearer96}  and  $HP(G_p)= (\sqrt {2p-1}+1)/2$ represents the upper bound on $\omega(G_p)$ established in \cite{HP21}.}
\label{fig:curves}
\end{figure}

 We replicated the $L^2(G_p)$ computations reported in \cite {GLV, Gvozd} using Matlab/CVX for primes  $p \leq 809$ as well as extended  them for all $p < 1000$.\footnote{The code is available at \url{http://vkobzar.com/}.}  These resulting new values  are shown in Table \ref{tab:l2}.  Figure \ref{fig:curves} shows that $L^2(G _p) \sim p^{0.456}$  which is tighter than $HP(G_p)= (\sqrt {2p-1}+1)/2$, the upper bound on $\omega(G_p)$ established in \cite{HP21,  DBSW21}.\footnote{For reference, we have also plotted and fitted to a power model the $L^3(G_p)$ values for  $p \leq 809$ determined in \cite {GLV, Gvozd}.} Moreover, since  $SOS_4(G_p) \leq L^2(G_p)$, our results provide some numerical evidence that the $SOS_4$ relaxations of the Paley graph clique number may be asymptotically growing at an order smaller than square root of $p$. However, due to the size of the SDPs, we have not been able to compute the $L^2(G_p)$ values for $p>997$. Therefore, our scaling estimate is not conclusive and presents an interesting open problem for further study. 
 \begin{table}
\begin{center}
 \begin{tabular}{||c c c c||} 
 \hline
 $p$     &  $\upvartheta (G_p)$ &  $L^2(G_p)$ &$\omega(G_p)$ \\ [0.5ex] 
 \hline
 821&28.653&18.673 & 12\\
829&28.792&18.105& 11\\
853&29.206&18.909  &13\\
857&29.275&18.429 &13\\
877&29.614&19.711 &13\\
881&29.682&18.689 &11\\
929&30.48&19.292  &13\\
937&30.61&19.248 &11\\
941&30.676&19.34 &11\\
953&30.871&19.199 & 11\\
977&31.257&19.737 &13\\
997&31.575&20.058 &13\\ [1ex] 
 \hline
\end{tabular}
\end{center}
\caption{The $L^2(G_p)$ values for $ 809< p < 1000$ determined in this paper, together with the values of $\omega(G_p)$  obtained from \cite{Shearer96}.}
\label{tab:l2}
\end{table}
 
 \section {Potential extensions}
 We hope that the $L^2(G_p)$ values can be either estimated analytically, or computed numerically for $p> 997$ leading to a new scaling estimate with a higher confidence level than that of the estimate obtained in this work. We also hope that the higher degree $L^t$ relaxations can be computed and used to upper bound the corresponding $SOS_{2t}$ values in the context of the Paley graph clique number and beyond.  To achieve new numerical results, it may be advantageous to decompose the $A_S(y)$ matrices in the positive semidefinite constraints in terms of  smaller matrices. A result of this kind was obtained in \cite{Passuello} with respect to the Lovasz $\upvartheta$ by decomposing the moment matrix in terms of the so-called ``zonal" matrices. Another possible alternative would entail adapting the approach used in \cite {MMP19} by computing a block diagonal relaxation with respect to a suitable local subgraph of a Paley graph.  We leave these potential extensions for future work.

\section{Acknowledgements}

\thanks{V.A.K. acknowledges support by NSF grant DMS-1937254. We thank Afonso Bandeira for suggesting the research direction that led to this paper;  we are also very grateful to him, Nebosja Gvozdenovic, Tim Kunisky, Monique Laurent and Chi Hoi Yip for their helpful input.}

\Urlmuskip=0mu plus 1mu\relax
\bibliography{mybib}{}
\bibliographystyle{plain}

\end{document}

%% file: macros.tex
\def \R {\mathbb{R}}

\theoremstyle{definition}